\documentclass[preprint2]{aastex}

\usepackage{epsfig}
\usepackage{graphicx}
\usepackage{graphics}

\def \etal   {\hbox{\it et~al.\/}}

\begin{document}

\shorttitle{Faraday Rotation in Wind-Blown Bubbles}

\title{Faraday Rotation Distributions from Stellar Magnetism in
Wind-Blown Bubbles
}

\author{R.~Ignace,}
\affil{
Department of Physics \& Astronomy, 
East Tennessee State University, Johnson City, TN, 37614, \\ ignace@etsu.edu}

\author{N.~M.~Pingel,}
\affil{
Department of Astronomy, University of Wisconsin-Madison, Madison, WI 53711,\\
nmpingle@wisc.edu}

\begin{abstract}

Faraday rotation is a valuable tool for detecting magnetic fields.
Here the technique is considered in relation to wind-blow bubbles.
In the context of spherical winds with azimuthal or split monopole
stellar magnetic field geometries, we derive maps of the distribution
of position angle (PA) rotation of linearly polarized radiation
across projected bubbles.  We show that the morphology of maps for
split monopole fields are distinct from those produced by the
toroidal field topology; however, the toroidal case is the one most
likely to be detectable because of its slower decline in field
strength with distance from the star.  We also consider the important
case of a bubble with a spherical sub-volume that is field-free to
approximate crudely a ``swept-up'' wind interaction between a fast
wind (or possibly a supernova ejecta shell) overtaking a slower
magnetized wind from a prior state of stellar evolution.  With an
azimuthal field, the resultant PA map displays two arc-like features
of opposite rotation measure, similar to observations of the supernova
remnant G296.5+10.0.  We illustrate how PA maps can be used to
disentangle Faraday rotation contributions made by the interstellar
medium versus the bubble.  Although our models involve simplifying
assumptions, their consideration leads to a number of general robust
conclusions for use in the analysis of radio mapping datasets.

\end{abstract}


\section{Introduction} 

Magnetism plays an important role in the lives of stars, frequently
in the form of its influence on angular momentum transport for star
formation and evolution (e.g., Ghosh \& Lamb 1979; Blandford \&
Payne 1982; Bodenheimer 1995; McKee \& Ostriker 2007; Meynet,
Eggenberger, \& Maeder 2011; Matt \etal\ 2012), and also in terms
of hot plasma generation (e.g., Davidson \& Ostriker 1973; Babel
\& Montemerle 1997ab; Townsend, Owocki, \& ud-Doula 2007; Li \etal\
2008; G\"{u}del \& Naz\'{e} 2009).
Of interest to this paper is the growing body of evidence for
magnetism among massive stars.  Direct detections of magnetism in
normal stars (i.e., not compact objects) are in large part relegated
to measuring circular polarizations in spectral lines arising from
the Zeeman effect (see Donati \& Landstreet 2009).  The first
detection of magnetism in a star besides the Sun dates back to
Babcock (1947).  Since then, the field has exploded.  A recent
review of the current state of the subject for non-degenerate stars
can be found in Donati \& Landstreet (2009).  The key result now
is that whereas magnetic detections were mainly limited to stars with
surface fields in the kilo-Gauss range, modern instrumentation and
diagnostic strategies regularly achieve highly significant
detections in the regime
of 100's of Gauss (e.g., Donati \& Collier Cameron 1997), and in
some special cases much less (e.g., Sennhauser \& Berdyugina 2011).

Numerous direct detections of surface magnetism in massive stars
have been reported, with some recent examples being Alecian \etal\ (2011),
Petit \etal\ (2011), Sch\"{o}ller \etal\ (2011), Hubrig \etal\ (2011, 2012)
Grunhut \etal\ (2012a, 2012b, 2012c), Wade \etal\ (2011; 2012).
These successes have correspondingly motivated theoretical
studies to understand the origin of these fields for massive stars
(e.g., MacGregor \& Cassinelli 2003; Braithwaite 2006; Cantiello
\etal\ 2009), their influence on massive star evolution (Maeder \&
Meynet 2003; Yoon, Dierks, \& Langer 2012), and connection to other
observational phenomena such as X-ray emissions (Babel \& Montmerle
1997b; Gagne \etal\ 1997; Gagne \etal\ 2005; Favata \etal\ 2009;
Ignace \etal\ 2010; Oskinova \etal\ 2011; Gagne \etal\ 2011; Wade
\etal\ 2012; Grunhut \etal\ 2012; Ignace, Oskinova, \& Massa 2012)
and aspherical wind flow (Poe, Friend, \& Cassinelli 1989; Shore
\& Brown 1990; Chevalier \& Luo 1994; ud-Doula \& Owocki 2002;
Townsend \& Owocki 2005; Brown, Cassinelli, \& Maheswaran 2008;
ud-Doula, Owocki, \& Townsend 2008; ud-Doula \etal\ 2012).

There are other diagnostics of stellar magnetism that have or could
complement the Zeeman-based approach.  Non-thermal radio emissions
from massive star colliding wind binaries have been used to infer
stellar magnetism (e.g., Williams \etal\ 1997; Dougherty \& Williams
2000; De Becker \etal\ 2006; van Loo, Runacres, \& Blomme 2006).
The Hanle effect is a weak Zeeman effect pertaining to the influence
of magnetic fields on the linear polarization in spectral lines.
The effect is sensitive to magnetic fields in the 1--100~G range
(depending on the Einstein A-value of the lines being measured) and
has been successfully used in studies of solar magnetism for decades
(e.g., Sahal-Brechot, Bommier, \& Leroy 1977; Stenflo 1982; Berdyugina
\& Fluri 2004; Trujillo Bueno \etal\ 2005).  There is a small but
growing literature on its potential application to other stars
(Ignace, Nordsieck, \& Cassinelli 1997; Lopez Ariste, Asensio Ramos,
Gonzalez Fernandez 2011; Ignace \etal\ 2011; Bommier 2012; Manso
Sainz \& Martinez Gonzalez 2012).

Another important method for measuring astrophysical magnetic fields
is Faraday rotation.  The effect refers to how the line-of-sight
(LOS) magnetic field component rotates the position angle of linear
polarization for a beam of radiation.  The amount of rotation is
also proportional both to the electron density (hence operates only
in a plasma) and to the path length through such regions.  Importantly,
the amount of position angle (PA) rotation scales with the square
of the wavelength of observation, $\lambda^2$.  In this way most
applications measure the polarization PA for a range of wavelengths
to derive a quantity called the ``rotation measure'' (or RM; see
the following section) that encodes information about the integrated
product of the LOS field component and electron number density.

Most applications of Faraday rotation are for interstellar or
extragalactic studies.  Space precludes a comprehensive review of
this literature; discussion of the state of the field, with references
therein, can be found for example in Carilli \& Taylor (2002), Han
\etal\ (2006), and Beck (2012).  Here attention is focused on the
potential of Faraday rotation as a probe of stellar magnetism in
wind-blown bubbles.  There has been several recent observational
developments that speak to the interaction of stellar winds or
supernova explosions with the surrounding interstellar environment
and the capacity of probing these interactions with Faraday rotation.

Ransom \etal\ (2008, 2010) have conducted studies of Faraday rotation
effects arising from planetary nebulae (PNe).  The relative motion
between a PN and the surrounding interstellar medium (ISM) alters the
strength and direction of the interstellar magnetic field, leading
to variations of the polarization PA across the PNe and the trailing tail
it creates in relation to the PAs surrounding the structure.  Although
implicitly placing constraints on the stellar magnetic field interior
to the nebula, the PNe act essentially as perturbers, on the scale
of a few parsecs, of the local medium, with Faraday rotation serving 
as a probe of the resultant disturbances.

Regarding massive star influences, Savage, Spangler, \& Fischer
(2012) report on an extensive study of Faraday rotation for the H{\sc ii}
region, the Rosette Nebula.  In this application a cluster of massive
stars lead to a wind-blown bubble and surrounding photoionized
region by the central OB stars.  In contrast to the PN studies of
Ransom \etal\ (2008, 2010), whose analysis was based on PA maps
with the diffuse Galactic synchrotron used as a source of linearly
polarized radiation, the Rosette study of Savage \etal\ analyzed
data from an array of sightlines to background extragalactic sources
of polarized emission that intercept the nebula and surrounding
region.  Their data are consistent with PA changes across the
nebula, on a scale of $\sim 10$~pcs, arising from the presence
of the bubble and its impact on the interstellar magnetic field.

For the case of Faraday rotation as a probe of stellar magnetism,
Harvey-Smith \etal\ (2010) discuss an antisymmetric RM morphology
across the supernova remnant (SNR) G296.5+10.0.  The SNR nebula has
two prominent emission arcs on opposite sides of a symmetry axis,
yet the PA rotations of either arc are oppositely oriented.  This
kind of pattern would be expected from a magnetic field that reverses
its LOS polarity from one arc to the other.  The authors associated
the pattern with the magnetic field of a slow magnetized red
supergiant wind.  The ejecta shell from the SN has ``swept up'' the
supergiant wind into the observed shell.  To reproduce the RM
pattern, the stellar magnetic field would have to be toroidal to
produce the observed polarity change between the emission arcs.
The authors derive an expression to relate the observed scale of
PA rotations to properties of the shell and the stellar wind.  The
observed change in amplitude of the RM by approximately 40 rad/m$^2$
indicates that a surface stellar magnetic field on the order of 500~G
could account for the observations.

In this paper we explore further the idea of Faraday rotation as a
means of deriving information about stellar magnetism in the
large-scale wind that might sometimes be observed in wind-blown
bubbles and SNRs.  In section~\ref{sec:model}, a brief review of
the expressions describing Faraday rotation are given.  A derivation
of PA rotation maps for an ionized and spherical stellar wind with
azimuthal (or toroidal) magnetic fields is presented in
section~\ref{sub:azi}.  PA rotation maps for a split monopole are
presented in section~\ref{sub:split} as a contrast case.  Returning
to the azimuthal field in section~\ref{sub:swept}, simple insertion
of a spherical subvolume taken to have no magnetic field is used
to simulate a two-wind interaction like that of a SNR or any scenario
in which a fast flow overtakes a slower one from an earlier stage
of stellar evolution.  In section~\ref{sec:disc} applications of
our results are discussed, with concluding remarks given in
section~\ref{sec:conc}.

\section{Theoretical Models}   \label{sec:model}

The standard expression to represent the polarization PA
rotation, $\psi$, arising from Faraday rotation along a sightline
is

\begin{equation}
\psi = \psi_0 + RM\,\times\left(\frac{\lambda}{1~{\rm m}}\right)^2,
	\label{eq:psi}
\end{equation}

\noindent where $\psi_0$ is the orientation of a background
polarization upon which Faraday rotation operates, and RM is
the ``rotation measure'' given by the path length integral of

\begin{equation}
RM = 0.81~{\rm rad~m^{-2}}\,\int\,\left(\frac{B_\parallel}{\rm \mu G}\right)\,
	\left(\frac{n_{\rm e}}{\rm cm^{-3}}\right)\,
	\left(\frac{dz}{\rm pc}\right),
	\label{eq:RM}
\end{equation}

\noindent where $B_\parallel$ is the LOS component of
the magnetic field, $n_{\rm e}$ is the electron density, and 
path length. 

The rationale for this particular formulation is that observers are
frequently interested in deriving RM from the radio data as a
constraint on the magnetic field strength (modulo the electron
density, whose value may be constrained from other considerations).
One measures values of $\psi$ for a range of wavelengths; plotting
$\psi$ against $\lambda^2$ in a log-log plot should then yield a
straight line whose slope is the rotation measure, RM.  If $RM>0$,
then the net Faraday rotation from all of the parallel field
components along the LOS, some being positive and some being negative,
in regions where there are electrons has resulted in a counter-clockwise
rotation of the background linearly polarized radiation.  If $RM<0$,
the net effect is a clockwise rotation.  The sign of $\psi$, or
alternatively RM, is sensitive to the electron-density averaged
net LOS field component along a given sightline.

The distinction being drawn here is that our models emphasize 
{\em differential} Faraday rotation (e.g., Minter \& Spangler 1996).
This amounts to a map of how RM varies between neighboring
sightlines, which is equivalent to analyzing maps of $\psi$ at a
fixed wavelength.  Consequently, it is convenient for our
purposes to reformulate
the effect of Faraday rotation as:

\begin{equation}
\psi = \frac{\pi}{z_0(\lambda)}\,\int\,\left(\frac{B_\parallel}{B_0}\right)\,
	\left(\frac{n_{\rm e}}{n_0}\right)\,dz
	\label{eq:psimap}
\end{equation}

\noindent where $n_0$ and $B_0$ are chosen normalizations,
$z_0(\lambda)$ is a relevant length scale for the problem, $dz$ is
in the same units of $z_0$, and the factor of $\pi$ is made explicit
here both to indicate that $\psi$ is measured in radians and to highlight the
fact that polarization PA is degenerate in multiples
of $\pi$ (or $180^\circ$).

It is straightforward to convert such a map in PA to one in RM.
If one evaluates $\psi(\lambda_0)$ at a reference wavelength $\lambda_0$,
then

\begin{equation}
\psi(\lambda) = \psi(\lambda_0)\times \frac{\lambda^2}{\lambda_0^2},
\end{equation}

\noindent and so

\begin{equation}
RM = \frac{\psi(\lambda_0)}{\lambda_0^2}.
\end{equation}

\noindent The following discussion of Faraday rotation 
in stellar bubbles adopts the use of equation~(\ref{eq:psimap}).

We focus on three illustrative cases to highlight expected PA map
morphologies as motivated by current observations.  First the case
of an azimuthal field distribution is considered.  Since the field
strength diminishes only as $r^{-1}$, it is this component that is
most likely to have observational relevance at the 1--10 parsec scales
of stellar bubbles.  Second, as a contrast case, PA maps are derived
for a split monopole field.  This case leads to results that are
morphologically distinct from the azimuthal field case.  The field
drops much faster with distance from the star, as $r^{-2}$, making
this case essentially unobservable; still, the split monopole has
value in providing insight into the range of morphological
possibilities.  Finally, we consider again the azimuthal field
scenario but now imposing a central ``cavity'' that does not
contribute to Faraday rotation (e.g., a central region devoid of a
magnetic field).  This approximates a ``swept up'' field from a
wind-wind interaction, such as may occur for a SNR
or a PN, where only the wind from an earlier phase
of stellar evolution has a relevant magnetic field.

We stress that spherical symmetry is adopted for the bubble shape
and density distribution.  It is not the goal of this paper
to reproduce the observations for any particular object.  A
spherically symmetric
density profile provides a ``controlled'' environment for
which to evaluate and gain insight into Faraday rotation through
stellar bubbles.  Real situations may involve a broad range of
additional (and potentially important) considerations, such as:
aspherical density distributions (e.g., bipolar flows or
clumping); radius-dependent
and/or aspherical
ionization effects; correlated behavior between density, ionization,
and/or a dynamically relevant magnetic field (e.g., how 
a toroidal field can lead to axisymmetric bubbles as
described in Chevalier \& Luo 1994).  Our spherical results 
provide a suite of baseline cases that are analytic or semi-analytic
in which one can evaluate modifications to predicted PA maps arising from
these more complex factors.

Before proceeding it is useful first to review the different
contributing components to Faraday rotation that affect
the final observed PA at the Earth.  The underlying assumption is
that Faraday rotation acts to rotate the PA of linearly polarized
radiation as it passes through magnetized and ionized regions.  The
background source (whether a point source or diffuse synchrotron background)
has some initial position angle $\psi_{\rm orig}$.  This value receives an
additional contribution $\psi_{ISM}$ owing to the 
ISM.  The total PA measured at the Earth is then

\begin{equation}
\psi_{\rm meas} = \psi_{\rm orig} + \psi_{ISM},
\end{equation}

\noindent where both $\psi_{\rm orig}$ and $\psi_{ISM}$ are independent
quantities that can vary from one sightline to the next.

Now for sightlines that intersect a stellar bubble, one must {\em
subtract} the ISM contribution for that segment of the sightline
passing through the bubble, which we signify as $\delta \psi_{ISM}$.
Then the contribution made by the bubble, $\psi_{\rm bub}$, must
then be {\em added}.  The total PA at the Earth now becomes:

\begin{equation}
\psi_{\rm meas} = \psi_{\rm orig} + \psi_{ISM} - \delta \psi_{ISM} +\psi_{\rm bub}.
\end{equation}

\noindent The calculations in the following sections present maps 
specifically of $\psi_{\rm
bub}$.  Section~\ref{sec:disc} deals with the fact that the bubble
contribution alone is not what is actually measured.

\subsection{Azimuthal Fields}	\label{sub:azi}

Modeling the PA morphologies of stellar bubbles requires the
introduction of two coordinate systems:  the observer system and
the stellar one.  For the observer Cartesian, cylindrical, and
spherical coordinates are adopted as $(x,y,z)$, $(\varpi, \alpha,
z)$, and $(r,\theta,\alpha)$, where the origin is the bubble center.
Here $z$ is the observer axis, with the Earth at $+\infty$.  The
angle $\alpha$ is measured counter-clockwise about this $z$-axis.
Then $\theta$ is the polar angle from the observer axis.  As we will
be discussing sightlines intersecting the stellar bubble, the
cylindrical radius $\varpi$ will be the impact parameter for such
rays.  For the star the corresponding Cartesian and spherical
coordinates are $(x_\ast, y_\ast, z_\ast)$ and $(r,\vartheta,\varphi)$.

For the transformation between these coordinates, we choose $y=y_\ast$.
Using unit vectors, the viewing inclination angle $i$
between the $z$ and $z_\ast$ axes is given by

\begin{equation}
\cos i = \hat{z}\cdot\hat{z_\ast}.
\end{equation}

\noindent Transformations between the angular quantities can be
obtained with spherical trigonometric relations, that will be used 
as needed.

In this first example, an azimuthal stellar magnetic field is
considered.  The vector field is $\vec{B} =  B_\varphi\,\hat{\varphi}$.
Calculation of Faraday rotation along a sightline requires determination
of the LOS field component, which is given by

\begin{equation}
B_\parallel = \vec{B}\cdot \hat{z} = B_\varphi\,(\hat{\varphi}\cdot\hat{z}).
\end{equation}

\noindent The transformation between the observer and stellar
coordinates is needed to evaluate the preceding dot product; the
rotation matrix between coordinate systems is given by

\begin{equation}
\left(\begin{array}{c} \hat{x_\ast} \\ \hat{y_\ast} \\ \hat{z_\ast} \end{array}
	\right) = \left( \begin{array}{ccc} \cos i & 0 & \sin i \\ 
	0 & 1 & 0 \\ -\sin i & 0 & \cos i \\ \end{array} \right) \,
\left(\begin{array}{c} \hat{x} \\ \hat{y} \\ \hat{z} \end{array} \right)\, .
\end{equation}

\noindent Consequently, one obtains

\begin{equation}
B_\parallel = -B_\varphi\,\sin \varphi\,\sin i.
\end{equation}

In addition to the LOS field component, the distribution of the
field strength with location about the star is also needed.  We
adopt the kinematic prescription put forth in Ignace, Bjorkman, \&
Cassinelli (1998) based on wind compression theory (Bjorkman \&
Cassinelli 1993).  The Ignace \etal\ model assumes a magnetic field
that is dynamically negligible as compared to the wind flow.  Assuming
flux freezing, the axisymmetric field topology can be derived,
a result they refer to as ``WCFields''.  Their model includes a
parameter for the axisymmetric density distribution.  For our
purposes this parameter, $d\mu/d\mu_0$, is set to unity for a
spherical wind; then using their asymptotic formula
(eq.~[21] of Ignace \etal\ 1998), the azimuthal
field becomes simply

\begin{equation}
B_\varphi(r,\vartheta) = B_{\rm eff} \,\frac{R_\ast}{r}\,\sin\vartheta,
\end{equation}

\noindent where $\vartheta$ is the co-latitude on the star, signifying
that the maximum toroidal field strength occurs at the rotational
equator of the star; and for the model of Ignace \etal, the
conveniently defined effective surface field strength, $B_{\rm
eff}$, depends on the actual surface field strength $B_\ast$, the
stellar rotation speed $v_{\rm rot}$, and the wind terminal speed
$v_\infty$, with

\begin{equation}
B_{\rm eff} = B_\ast\,\left(\frac{v_{\rm rot}}{v_\infty}\right).
\end{equation}

Calculation of the Faraday rotation proceeds from a sightline-dependent
integration through the bubble, as given by

\begin{equation}
\psi_{\rm bub}(\varpi, \alpha) = \frac{\pi}{z_{\rm b}}\,\int\,\left[\frac{B_\varphi
	(r,\vartheta)}
	{B_{\rm eff}}\right]\,\left[\frac{n_{\rm e}(r)}{n_{\rm w}}\right]\,(\hat{z}
	\cdot\hat{\varphi})\,dz,
\end{equation}

\noindent with $n_{\rm e} = n_{\rm w}\,(R_\ast/r)^2$, where $n_{\rm
w}$ is a density scale associated with the wind.  This integral
reduces to

\begin{equation}
\psi_{\rm bub}(\varpi, \alpha) = -\frac{\pi}{z_{\rm b}}\,\int\,\left(\frac{R_\ast}
        {r}\right)^3\,\sin\vartheta\,\sin\varphi\,\sin i\, dz.
\end{equation}

\noindent Note that $\sin\vartheta\sin\varphi \equiv y_\ast/r =
y/r$ for our coordinate system definitions.  Given that
$y\perp z$, the coordinate $y$
can be factored out of the integral which now becomes

\begin{equation}
\psi_{\rm bub}(\varpi, \alpha) = -\pi\,\sin i\, \frac{y}{z_{\rm b}}\,\int\,
	\left(\frac{R_\ast}{r}\right)^4\,\frac{dz}{R_\ast}.
\end{equation}

\begin{figure}[t!]
\plotone{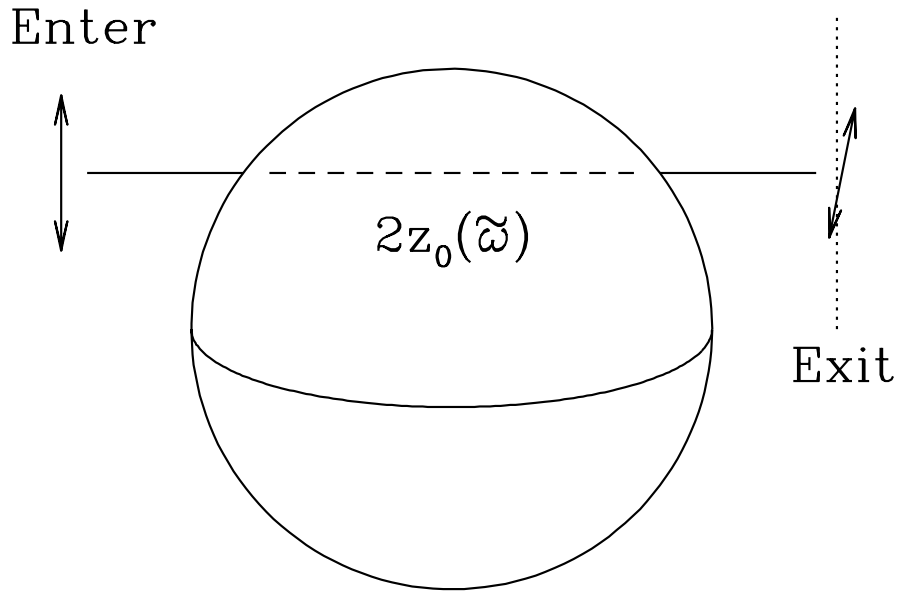}
\caption{Illustration of the geometry for evaluating 
PA changes across spherical bubbles because of
Faraday rotation.  The observer is located off to the right.
Linearly polarized radiation moving left to right passes
through the bubble.  Along an arbitrary sightline of
impact parameter $\varpi$, the path
length through the bubble will be given by $z_0-(-z_0) 
= 2z_0=2\sqrt{R^2-\varpi^2}$, where $R$ is the radius
of the bubble.  Upon emerging, the orientation of
the linear polarization against the sky will have rotated
in a manner that depends on the magnetic field and electron density
along the path.
\label{fig1}}
\end{figure}

\noindent To solve this equation, we note that $r^2=\varpi^2+z^2$, where
$\varpi$ is the impact parameter of the sightline under consideration,
and therefore a constant of the integration.  The sightline enters
the bubble at $+z_0$ and exits at $-z_0$, where
$z_0(\varpi) = \sqrt{R^2-\varpi^2}$ (see Fig.~\ref{fig1}).
Back-front symmetry of the integration gives 

\begin{equation}
\psi_{\rm bub}(\varpi, \alpha) = -2\pi\,\sin i\, \frac{y}{z_{\rm b}}\,
	\int^{z_0(\varpi)}_0\,\left(\frac{R_\ast^2}{\varpi^2+z^2}\right)^2
	\,\frac{dz}{R_\ast},
\end{equation}

\noindent which, after some rearrangement, has the solution

\begin{eqnarray}
\psi_{\rm bub}(x,y) & = & -2\pi\,\left(\frac{R_\ast}{z_{\rm b}}
	\right)\,\left(\frac{y}{\varpi}\right)\,\left(\frac{R_\ast}{\varpi}
	\right)^2\,\sin i \nonumber \\
 & & \times \left[(\pi/4 - \theta_0/2) + \frac{1}{4}\sin 2\theta_0\right],
	\label{eq:azsoln}
\end{eqnarray}

\noindent where $\tan \theta_0 = \varpi/z_0(\varpi)$ and $\varpi=
\sqrt{x^2+y^2}$.

There are several key comments to be made about this solution.

\begin{itemize}

\item First, the appearance of the factor of $y$ means
that the PA map is left-right {\em antisymmetric} about
the line of $x=0$ in the plane of the sky.

\item Second, the overall morphology of the PA map is {\em independent} of
viewing inclination.  The inclination angle appears in the solution
only in the multiplicative factor $\sin i$, acting as an
amplitude scale.  Consequently, the PA map for the edge-on view is the
same map that results for any other inclination, just the amount
of PA rotation is reduced for every sightline by $\sin i$.

\end{itemize}

\begin{figure}[t!]
\centering{\includegraphics[width=3.9in]{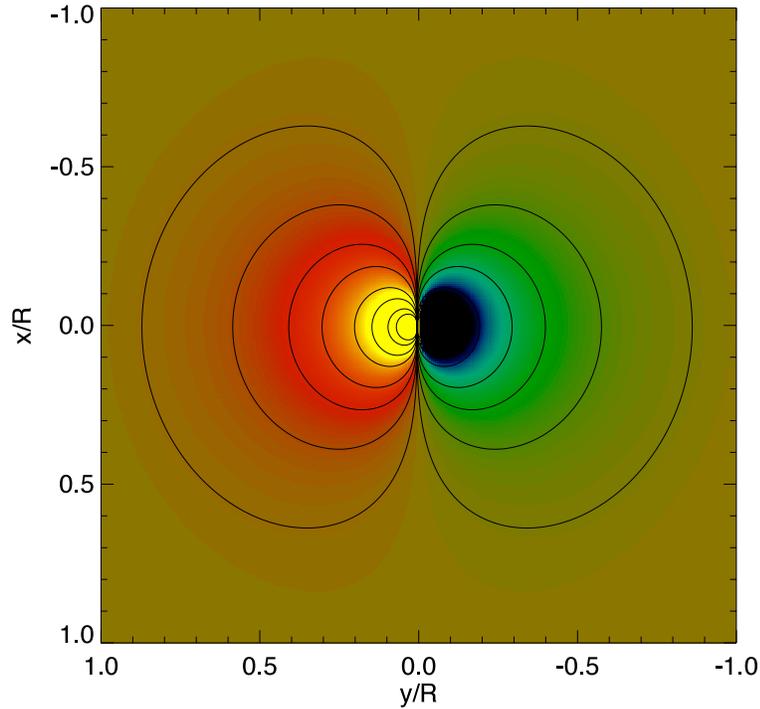}}
\caption{A false color image with contours overlaid indicating
the amplitude of the PA changes that would
be observed across a wind-blown bubble with an azimuthal
magnetic field.  The plot is for
the bubble contribution only.  The green and blue colors
are negative PA rotations; yellow and red are for positive values.
The plot space is for the plane of the sky as $y$ versus $x$,
with the coordinates normalized to the radius of the bubble
$R$.  Note the antisymmetric morphology, and that $PA=0^\circ$
along the $y=0$ axis.  Further, this morphology is valid for
{\em all} viewing inclinations, with the amplitude of the PA
rotation scaling linearly with $\sin i$.
\label{fig2}}
\end{figure}

Figure~\ref{fig2} shows a false-color plot of the PA map (or after
proper normalization, the RM map) for the solution of
equation~(\ref{eq:azsoln}).  Based on our conventions, the positive
$x$-axis points down in this figure, and the positive $y$-axis
points right; together these give positive $z$ toward the observer.
The map is indeed left-right antisymmetric and top-down symmetric.
Additionally, because the azimuthal field is everywhere perpendicular
to the LOS for $x=0$, $\psi_{\rm bub}=0$ along the vertical that
passes through the map center.

No absolute scale is given in this figure as it depends on a number
of wind and star parameters, such as the surface field strength,
the mass-loss rate, wind speed, and stellar radius as well as the
bubble radius (hence, its age).  We return to the expected level
of PA and applications in section~\ref{sec:disc}.

\subsection{Split Monopole Fields}	\label{sub:split}

It is useful to consider a different field topology to explore the
range of PA map morphologies that can result in stellar bubbles.
An azimuthal field exhibits the most shallow radial decline expected for
a field that is carried out by a stellar wind.  The next most shallow
decline would be a split monopole.  This is a radial magnetic field
with a field strength $B_{\rm r} = B_\ast\,(R_\ast/r)^2$, but the
polarity changes sign from one hemisphere to the other.  So the
field is outward directed in one hemisphere, but inward in the
other.  Such a magnetic configuration is what would be expected of
a strong stellar wind that distorts a dipole field at the star
into a radial geometry (e.g., ud-Doula \& Owocki 2002).

Calculation of PA maps proceeds as before, except that now there
are different factors appearing in the integrand because of
the new field topology.  Instead of
$\hat{z}\cdot\hat{\varphi}$ for the azimuthal field case, we now
have $\hat{z}\cdot\hat{r}=z/r$ for a radial field.
The integral takes on the form:

\begin{equation}
\psi_{\rm bub} = - \frac{\pi}{z_{\rm bub}} \, \int\, 
	\left[\frac{n_{\rm e}(r)}{n_{\rm w}}\right]
	\left[\frac{B_{\rm r}(r)}{B_\ast}\right]\, \left(\frac{z}{r}\right)\,dz.
\end{equation}

Before proceeding, the split monopole case offers a new wrinkle for
the calculation of the Faraday rotation.  The radial field switches
polarity between hemispheres.  Consequently, at a general viewing
inclination, the net Faraday rotation cancels identically for
sightlines which do {\em not} intercept the magnetic equator.  Such
sightlines enter and exit the bubble in a hemispherical cap of just
one field polarity.  As a result, there is as much
clockwise PA rotation through, say, the first half of the pathlength
as there is counterclockwise contribution through the second half.
The radial field ensures that for every value of $B_\parallel$ along
the path, there is a corresponding value of $-B_\parallel$ at a
reflected position in a back-front sense along the path.  These two
positions of opposed LOS field components occur at the same radius
from the star, and therefore occur at the same density.  A polarity
switch, and therefore a net Faraday rotation, only occurs for
sightlines that intercept the magnetic equator.

The result of all this is that a pair of truncated hemispherical
caps appear at top and bottom in the PA maps for the split monopole
case.  The extent of these truncations depends on the viewing
inclination.  For the pole-on case, every sightline passes through
the magnetic equator, and there are no truncation zones.  The extreme
opposite case is the edge-on view; here, no sightlines pass through
the magnetic equator, and so $\psi_{\rm bub}$ is identically zero
everywhere.  In observer coordinates the truncation occurs for $x
\ge \pm R\cos i$.

For sightlines that do intercept the magnetic equator, the location
along the path where this occurs, $z_{\rm eq}$, must be known so
that the sign change is properly taken account in the integration.
For sightlines with $|x| < R \cos i$, the integral becomes

\begin{equation}
\psi_{\rm bub} = -\frac{\pi}{z_{\rm b}}\,R_\ast^4\,\left\{
	\int_{z_{\rm eq}}^{+z_0(\varpi)}\,\frac{z}{r^5}\,dz
	-\int_{-z_0(\varpi)}^{z_{\rm eq}}\,\frac{z}{r^5}\,dz
	\right\},
	\label{eq:psisplit}
\end{equation}

\noindent where a sightline with impact
parameter $\varpi$ at observer azimuth $\alpha$ for a split
monopole viewed at inclination $i$ will intercept the equatorial
plane at location

\begin{equation}
z_{\rm eq} = -\varpi\,\tan i\,\cos\alpha.
\end{equation}

\noindent Evaluating the integral of equation~(\ref{eq:psisplit}),
along with some algebraic manipulation, leads to

\begin{equation}
\psi_{\rm bub} = -\frac{2\pi}{3}\,\left(\frac{R_\ast}{z_{\rm b}}\right)\,
	\left(\frac{R_\ast}{R}\right)^2\left[\left(\frac{R^2\,\cos^2 i}
	{x^2+y^2\cos^2 i} \right) - 1\right].
\end{equation}

\begin{figure}[t!]
\centering{\includegraphics[width=3.9in]{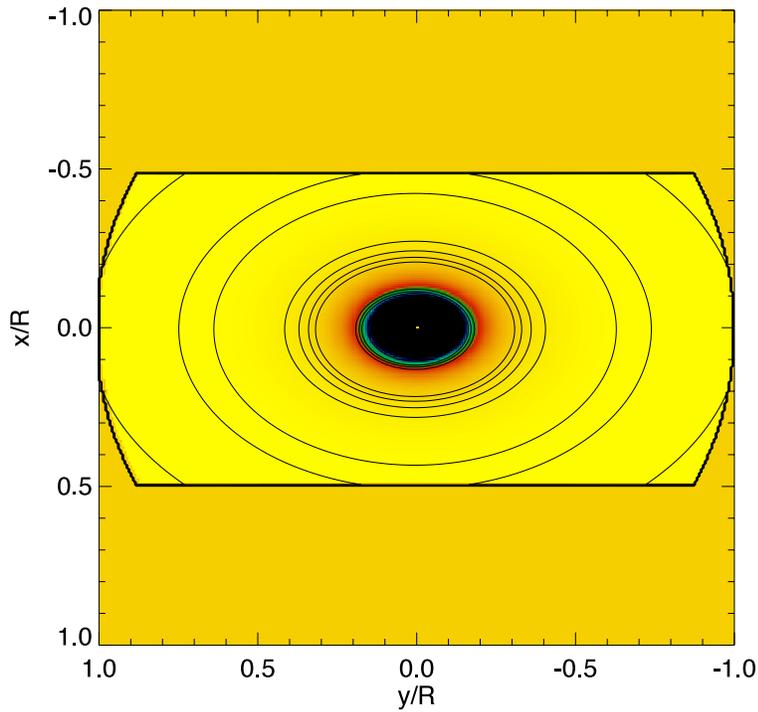}}
\caption{Similar to Fig.~\ref{fig2}, this PA map is for a split
monopole field.  In contrast to the azimuthal field case, the PA
rotations are all of the same sense, either clockwise or counterclockwise
(depending on the polarity of the field axis directed toward the observer).  
Sightlines that fail to intercept the magnetic equator when
$|x|>R\cos i$ (see text)
make
no contribution to a net Faraday rotation, leading to the appearance
of ``truncated'' hemispheres; the example shown
is for $i=60^\circ$.
\label{fig3}}
\end{figure}

\noindent This solution is displayed in Figure~\ref{fig3} at
a viewing inclination angle of $i=60^circ$.

Unlike the case of an azimuthal field, the scaling of the results
for the PA is more complicated with viewing inclination.  The
truncated caps are inclination dependent.  The PA maps for a split
monopole are markedly different from that of an azimuthal field as
seen in Figure~\ref{fig2}.  Whereas the azimuthal field produces
an antisymmetric morphology, all the Faraday rotations for a split
monopole are of the same sense:  either everywhere clockwise or
everywhere counterclockwise.

As noted previously, the effects of Faraday rotation are much smaller
across the bubble for a split monopole than for an azimuthal
field.  However, assuming that $\psi_{\rm bub} \approx 0$ for the
vast majority of sightlines in the split monopole case, the bubble
still influences the PA map of the region because of differential
RM effects for sightlines intercepting the bubble versus those that
do not.  In other words even if $\psi_{\rm bub} \ll 1$, the mere
presence of the bubble means that $\delta \psi_{ISM}$ may still be
significant.  More will be said about this in section~\ref{sec:disc}.

\subsection{``Swept Up'' Fields}	\label{sub:swept}

For the last application, consider again the azimuthal field case.
To simulate a two-wind interaction, where an inner wind (or a
supernova ejecta shell) is blown into, and sweeps up, an outer wind
from a earlier stage of stellar evolution, we impose a central
spherical region that does not contribute to Faraday rotation.  This
could be either because the central region has no magnetic field
or because it consists of neutral gas (the latter being unlikely
for applications to massive stars).  The overall scenario is intended
to approximate situations like the production of a PN, or a fast
wind from a blue supergiant overtaking a slower denser red giant
wind (e.g., Chita \etal\ 2008), or a SN explosion that expands into
the wind of a progenitor phase.

To be specific, we define an interior boundary
of radius $R_1 < R_2$, now with $R_2$ the outer bubble radius.  For
$r<R_1$ the region is effectively a ``cavity'' in terms of its
contribution to the Faraday rotation along a sightline.  The 
polarization PA, $\psi_{\rm bub}$, now becomes

\begin{equation}
\psi_{\rm bub} \propto y\, \sin i\times \int_{z_1}^{z_2}\,dz/r^3,
\end{equation}

\noindent where $z_1 = 0$ for $\varpi \ge R_1$, but for $\varpi < R_1$,
$z_1$ takes on the value

\begin{equation}
z_1 = \sqrt{R_1^2-\varpi^2}.
\end{equation}

\noindent Detailed steps for the derivation of $\psi_{\rm bub}$
as a function of sightline, along with a generalization to a
power-law density, are given in the Appendix.
Figure~\ref{fig4} shows examples of the resulting
PA maps for cavities of different relative extents with $R_1/R_2
=0.25$, 0.5, and 0.75.  The maps remain left-right antisymmetric
as in section~\ref{sub:azi},
but now two maxima appear in the form of arcs.  

Not surprisingly, these maps are quite reminiscent of the morphologies
seen in SNR G296.5+10.0 (Harvey-Smith \etal\ 2010).  In that report
a discussion and derivation similar to this paper are presented.
Harvey-Smith \etal\ introduced a toroidal field to explain the
antisymmetric dependence of RM observed in the SNR.  Those authors
additionally included density and magnetic field enhancements arising
from a shock.  These are additional relevant ingredients that we
have not incorporated here, but such effects could be included either
in a phenomenological way or using detailed dynamical simulations.
Our approach highlights the robust nature of the
antisymmetric PA map morphology provided by an azimuthal field along
with the appearance of arc-like features arising from the presence
of a ``cavity'', a region devoid of any contribution to Faraday
rotation.

\section{Discussion}	\label{sec:disc}

The objective of this paper has been to elucidate the
effects of Faraday rotation through a stellar bubble by evaluating
of PA maps (or equivalently RM maps) under certain simplifying
assumptions to focus on broad morphological trends.  As such,
spherical symmetry has been assumed for the geometry of the bubble
and the run of density, specifically a wind type density that drops
as $r^{-2}$ with radius.  At this point it is worth commenting on
the limitations of this approach, the kinds of effects that need
to be included for application to particular types of bubbles, and
strategies for best extracting information about the properties of
the bubble and its immediate environment.

The case of an azimuthal stellar magnetic field is likely
the only one of observational relevance because of its
more gradual decline in field strength as $r^{-1}$ to allow 
for detection at the 1--10~pc scale of stellar bubbles.  A
robust prediction of even the simplified models presented
here is that a sign reversal in polarization PA
(or RM) is expected when an azimuthal field
plays a role in the Faraday rotation.  However, that sign 
reversal will only be achieved if the $\psi_{\rm bub}$ term
can be isolated.

Imagine a situation in which the ISM along a sightline passing near to
a bubble, but without intercepting the bubble, produces a PA value
of the form:

\begin{equation}
\psi_{ISM} = \pi\times L_0/z_{ISM},
\end{equation}

\noindent where $L_0$ is the relevant length scale over
which one can define an average product of the interstellar magnetic field
and electron number density. In other words:

\begin{equation}
\langle B_\parallel\,n_{\rm e}\rangle = \frac{1}{L_0}\,\int_0^{L_0}\,
	B_\parallel\,n_{\rm e}\,dz.
\end{equation}

\noindent Assuming that this average value is what would have been
sampled for a sightline through the bubble, and further assuming spherical
symmetry, the decrement to the interstellar PA rotation caused by
the presence of the bubble as a function of impact parameter is

\begin{equation}
\delta \psi_{ISM}(\varpi) = \psi_{ISM}\,\frac{2R}{L_0}\,\sqrt{1-\frac{\varpi^2}{R^2}}.
\end{equation}

The ISM is highly inhomogeneous in terms of its magnetic field
distribution, density, and ionization.  Consequently, one
hardly expects that $n_{\rm e}B_\parallel$ at the location of the
wind bubble would equal the pathlength-averaged value $\langle
B_\parallel\,n_{\rm e}\rangle$.  Still, 
it is useful to consider the resultant PA map from the combination
of the different contributions by adopting such a scenario.  
Choosing $\psi_{ISM}\footnote{Recall that polarization PA is
degenerate in multiples of $180^\circ$; here $\psi_{ISM}=1000^\circ$
amounts to a PA of $100^\circ$, as seen in Fig.~\ref{fig5}b.} = 1000^\circ$
and $2R/L_0 = 0.01$, Figure~\ref{fig5}a shows a PA map and contours
for an azimuthal field case, similar to the style of previous
figures, but now with the Faraday rotation contribution from the ISM included.  
Additionally, Figure~\ref{fig5}b displays what the measured
PA would be along an axis at $x=0.1R$.

Figure~\ref{fig5} is important for illustrating how to infer the
various components to the measured value of $\psi_{\rm meas}$.  If one could
measure the PA map across a wind-blown bubble with an azimuthal
field, the following outlines how to decompose the trace
of the signal in Figure~\ref{fig5}b to determine 
$\psi_{ISM}$, $\delta \psi_{ISM}$, and $\psi_{\rm bub}$.
Note that it is assumed that the interstellar field and
electron density in the vicinity of the bubble is approximately
constant (or at least smoothly varying).

\begin{itemize}

\item For $|y| > R$, the PA is set by sightlines that do not
intercept the bubble and therefore sample the interstellar
field and electron density.

\item For $|y|<R$, the sightlines intercept the bubble. 
The azimuthal field leads to $\psi_{\rm bub}$
that is predicted to be antisymmetric, but the decrement term
$\delta \psi_{ISM}$ is symmetric.  The result is a left-right asymmetric
profile.

\item The two peaks in Figure~\ref{fig5}(b) can be used to deduce
the amplitude of $\psi_{\rm bub}$.  The vertical blue double-arrowed
line represents the total change in the PA rotation owing to the bubble
contribution.  Introducing $\psi_{\rm max}(x) = |\psi_{\rm bub}(x,y_{\rm
max})|$ as being the maximum PA rotation through the bubble at
location $(x,\pm y_{\rm max})$, the blue arrow has an amplitude of
$2\psi_{\rm max}$.  The value of $\psi_{\rm max}$ is related to the
scale length $z_{\rm b}$ that depends on the wavelength of observation,
the density scale of the bubble, and importantly the surface field strength
of the star.  Bisection of the vertical arrow, as illustrated with
the horizontal dotted line, yields the value of $\psi_{\rm max}$.

\item The bisection mentioned in the previous point
gives the maximum decrement due to $\delta \psi_{ISM}$.  
With spherical symmetry, one should
expect a distribution of this decrement as indicated by the
green dashed line.  Bear in mind that the particular example
of Figure~\ref{fig5} implicitly assumes a local interstellar
field of positive polarity, which leads to the bowl-shaped
decrement as shown.  If the local field were of the opposite
polarity, the ``decrement'' would in fact produce an {\em 
inverted} bowl shape, meaning that the green line for
$\delta \psi_{ISM}$ would lie above $\psi_{ISM}$.  Consequently,
the decomposition process gives both the strength of the
local magnetic field and its polarity.  As one last comment,
it is possible that no decrement is found.  Such a result
would arise if the interstellar field in the locale of the bubble
were very low, or if the local ISM were of low ionization.

\end{itemize}

To be yet more quantitative, it is possible to 
relate a measured value of $\psi_{\rm max}$ from
the bubble contribution to the length scale $z_{\rm b}$.
Equation~(\ref{eq:azsoln}) gives the solution for $\psi_{\rm bub}$
in the azimuthal field case.  Figure~\ref{fig6} plots the value
of $|y_{\rm max}|$ as a function of $x$.  In other words
holding $x$ fixed, and measuring the polarization PA in a direction
perpendicular to the axis of antisymmetry, Figure~\ref{fig6}
provides the location $(x,y_{\rm max})$.  

The scaling for the PA rotation increases rapidly with
decreasing value of the impact parameter, $\varpi$.  Consequently,
the strongest measurable effect of the bubble for Faraday rotation
will be around the center of the projected bubble.  At these
locations one has $\theta_0\approx 0$, for which $y_{\rm max}$
is easily derived to be

\begin{equation}
y_{\rm max} \approx x/\sqrt{2},
\end{equation}

\noindent which is indicated by the blue line in Figure~\ref{fig6}.
The value of $\psi_{\rm max}$ is also straightforwardly
derivable as

\begin{equation}
\psi_{\rm max} \approx \sqrt{\frac{1}{27}}\,\pi^2\,\sin i\,
	\left(\frac{R_\ast^3}{z_{\rm b}\,x^2}\right).
\end{equation}

\noindent where $\psi_{\rm max}$\footnote{The reader might be
concerned that the value of $\psi_{\rm max}$ appears to diverge as
$x\rightarrow 0$, for which $y_{\rm max} \rightarrow 0$ as well.  First,
the apparent divergence does not take place owing to the finite size
of the star.  Second, although large values of $\psi_{\rm max}$ can
be achieved, the polarization PA suffers a $180^\circ$ degeneracy.
Many rotations in $\psi$ over a small observed angular scale rapidly
leads to polarimetric cancellation.  For expected star, wind, and
bubble parameters, the vast majority of the bubble will like have
at best a modest value of $\psi$, except perhaps at quite long
wavelengths.} is measured from the data corresponding to a selection
of $x$ on the PA map.  From these measures one can determine $z_{\rm
b}/\sin i$ from the data.  In the absence of any other information,
there is an inclination ambiguity.  Still, an upper limit to
$z_{\rm b}$ is obtained from assuming an edge-on view of the system,
in which case a lower limit to the stellar surface magnetic field
strength is also obtained.

The overall scale of the PA rotation through a stellar bubble is
set by a combination of stellar and wind parameters and the size
of the bubble itself (which, of course, is related to its age).
For the azimuthal field case, which is the one most likely to be
of observational significance, we have already that $B_\varphi =
B_{\rm eff}\,(R_\ast/r)$.  For the number density of electrons in a
spherical stellar wind, we have been using $n_{\rm e} = n_{\rm w}\,
(R_\ast/r)^2$.  The density scale constant is given by

\begin{equation}
n_{\rm w} = \frac{\dot{M}}{4\pi\,\mu_{\rm e}m_H\,v_\infty\,R_\ast^2},
\end{equation}

\noindent where $\dot{M}$ is the mass-loss rate, $\mu_{\rm e}$ is
the mean molecular weight per free electron, and $v_\infty$ is the
wind terminal speed.  Then the scale constant for Faraday rotation
becomes

\begin{eqnarray}
\psi_0  & = & 6.0^\circ\times\frac{\left(\frac{B_{\rm eff}}{10~{\rm G}}\right)\,
	\left(\frac{R_\ast}{
	100~R_\odot}\right)\,\left(\frac{\dot{M}/\mu_{\rm e}}
	{10^{-6}~M_\odot~{\rm yr}^{-1}}\right)}
	{\left(\frac{v_\infty}{100~{\rm km~s}^{-1}}\right)\,
	\left(\frac{R}{1~{\rm pc}}\right)^2} \nonumber \\
 & & \times \left(\frac{\lambda}{30~{\rm cm}}\right)^2.
\end{eqnarray}

\noindent This scale parameter is found to be larger for stars
with stronger surface fields, larger size, and higher mass-loss
winds; it is smaller for faster winds and larger (older) bubbles.
It also has the standard quadratic dependence on wavelength.

Using the preceding expressions, one can derive a ratio of
the stellar radius to the Faraday rotation length scale for
the bubble, $z_{\rm b}$, to be

\begin{eqnarray}
\frac{R_\ast}{z_{\rm b}} & = & 4\times 10^{11}\times
	\frac{\left(\frac{B_{\rm eff}}{10~{\rm G}}\right)\,
	\left(\frac{\dot{M}/\mu_{\rm e}}
	{10^{-6}~M_\odot~{\rm yr}^{-1}}\right)}
	{\left(\frac{v_\infty}{100~{\rm km~s}^{-1}}\right)\,
	\left(\frac{R_\ast}{
	100~R_\odot}\right)} \nonumber \\
 & & \times \left(\frac{\lambda}{30~{\rm cm}}\right)^2.
	\label{eq:zb}
\end{eqnarray}

\noindent The scale of this ratio is huge and thus warrants a comment
of interpretation.  Implicit is that the length scale for Faraday
rotation $z_{\rm b}$ has been evaluated for the magnetic field and
wind density near the stellar surface, which are enormous compared
to interstellar conditions.  As a result, $z_{\rm b}$ is driven to
incredibly small values on the order of 1~cm and less.  However,
expected values of Faraday rotation across the bubble are much
smaller than at the scale of the star.  Defining an associated
scale $z'_{\rm b}$ related to the dimensions of the bubble instead
of the star gives $(R/z'_{\rm
b}) = (R_\ast/z_{\rm b})\times (R_\ast/R)^2$, which when using the
parameterization of equation~(\ref{eq:zb}) is of order $10^{-2}$.
In other words the length scale associated with rotation of 
the polarization PA through $180^\circ$ is roughly 100 times larger
than the radius of the bubble.

\section{Conclusions}	\label{sec:conc}

The goal of this study has been to develop insight into the
possibilities of using Faraday rotation to probe stellar magnetism
in wind-blown bubbles.  Adopting spherical symmetry and considering
two field topologies of azimuthal and split monopole fields,
analytic and semi-analytic results were presented in the form of
PA maps arising from Faraday rotation across a stellar wind bubble.
The field strength for a split monopole declines too fast to have
detectable effects for Faraday rotation.  Instead, only the azimuthal
field scenario is likely to lead to detectable signals, for which the key
result derived here is the antisymmetric morphology for the PA
rotations (or equivalently, the RMs) across the bubble in addition
to a simple $\sin i$ scaling of the amplitude of PA rotation.
Inclusion of a central spherical region of zero magnetic field leads
to prominent and antisymmetric PA rotations in arc-like structures
like those observed in SNR G296.5+10.0 (Harvey-Smith \etal\ 2010).

To date, studies of Faraday rotation effects across H{\sc ii} regions
appear consistent with variations in the interstellar magnetic field
(e.g., Harvey-Smith, Madsen, \& Gaensler 2011).  Clearly there may
be only a restricted subset of bubbles for
which our models will have applications.
Moreover, dispersion in the interstellar RM may present challenges
in detection and interpretation of the effects we have described
even when present, as for example variations in the interstellar RM can
have typical amplitudes of order 10 rad/m$^2$ (Mao \etal\ 2010).
We are considering strategies for interpreting the interaction of
interstellar magnetic fields with stellar bubbles using Faraday
rotation effects.  Given recent results in relation to SNR G296.5+10.0,
the PN cases (Ransom \etal\ 2008, 2010), and the Rosette Nebula
(Savage \etal\ 2012), Faraday rotation offers great promise as a means
of using stellar bubbles to discern the properties of stellar and/or
interstellar magnetic fields.

\acknowledgments 

The authors are grateful to Steve Gibson for many helpful discussions
on astrophysical problems involving Faraday rotation and also to
an anonymous referee for comments that have improved this paper.
Ignace acknowledges support for this research through a grant from
the National Science Foundation (AST-0807664).  Pingel acknowledges
support from the National Science Foundation Research Experiences for
Undergraduates (REU) program through grant AST-1004872.

\appendix

\section{Appendix:  Generalized Solution for a Stellar Wind with
an Azimuthal Magnetic Field}	\label{app}

For an azimuthal magnetic field of the form $B_\varphi \propto 1/r$,
there is a general solution to the PA distribution arising from
Faraday rotation for a spherically symmetric envelope with an
electron density that is a power-law in radius, $n_{\rm e} \propto
r^{-m}$, with $m$ the power-law exponent.  Consider a bubble of
outer radius $R_2$ and inner radius $R_1$.  For $r<R_1$, the
interior region makes no contribution to the Faraday rotation.
Using $\theta$ as the variable of integration, with $\tan \theta
=\varpi/z$, a sightline enters the outer bubble edge at $\theta_2$
and enters the inner region at $\theta_1$.  Then the 
solution for the PA rotation arising from the bubble contribution is

\begin{equation}
\psi_{\rm bub} = \pi\,\left(\frac{2R_2}{z_{\rm b}}\right)\,\left(\frac{R_2}{\varpi}
	\right)^m\,\left(\frac{y}{\varpi}\right)\,\sin i\,
	\int^{\theta_1}_{\theta_2}\,\sin^m \theta\,d\theta.
	\label{eq:azigen}
\end{equation}

\noindent Although the preceding integral is mathematically valid,
it may not be physically plausible for arbitrary values of $m$.
The selection of $B_\varphi \propto r^{-1}$ essentially assumes a
frozen-in field that is dragged out with the wind plasma in a
constant expansion flow (e.g., Weber \& Davis 1967), implying an
inverse square-law density.  However, $m\ne 2$ could represent
changes in the ionization state of the gas with distance from the
star.

The integration of eqn~(\ref{eq:azigen}) takes into account
the front-back symmetry of the situation by integrating only over
the front hemisphere facing the observer.  This gives rise to the
factor of 2 appearing in the first fraction.  The integration limits
also allow for an interior central cavity of zero field (or,
alternatively, zero ionization) to the wind-blown bubble as was
discussed in the swept-up wind case of section~\ref{sub:swept}.

With or without a central cavity, the upper limit to the integral
is always given by

\begin{equation}
\sin \theta_2 = \varpi/R_2,
\end{equation}

\noindent for a bubble of outer radius $R_2$.  The presence of a
cavity only affects the lower limit to the integral.  If there
is no cavity, then $\theta_1 = \pi/2$ for all sightlines.
However, within an interior cavity of radius $R_1$, the lower
limit of $\theta_1 = \pi/2$ holds only for sightlines
that {\em fail} to intercept the cavity region, with $\varpi
\ge R_1$.  Sightlines with $\varpi<R_1$ pass
through the cavity; the upper limit to the integration now becomes

\begin{equation}
\sin \theta_1 = \varpi/R_1.
\end{equation}

\begin{figure}[t!]
\centering{\includegraphics[width=3.3in]{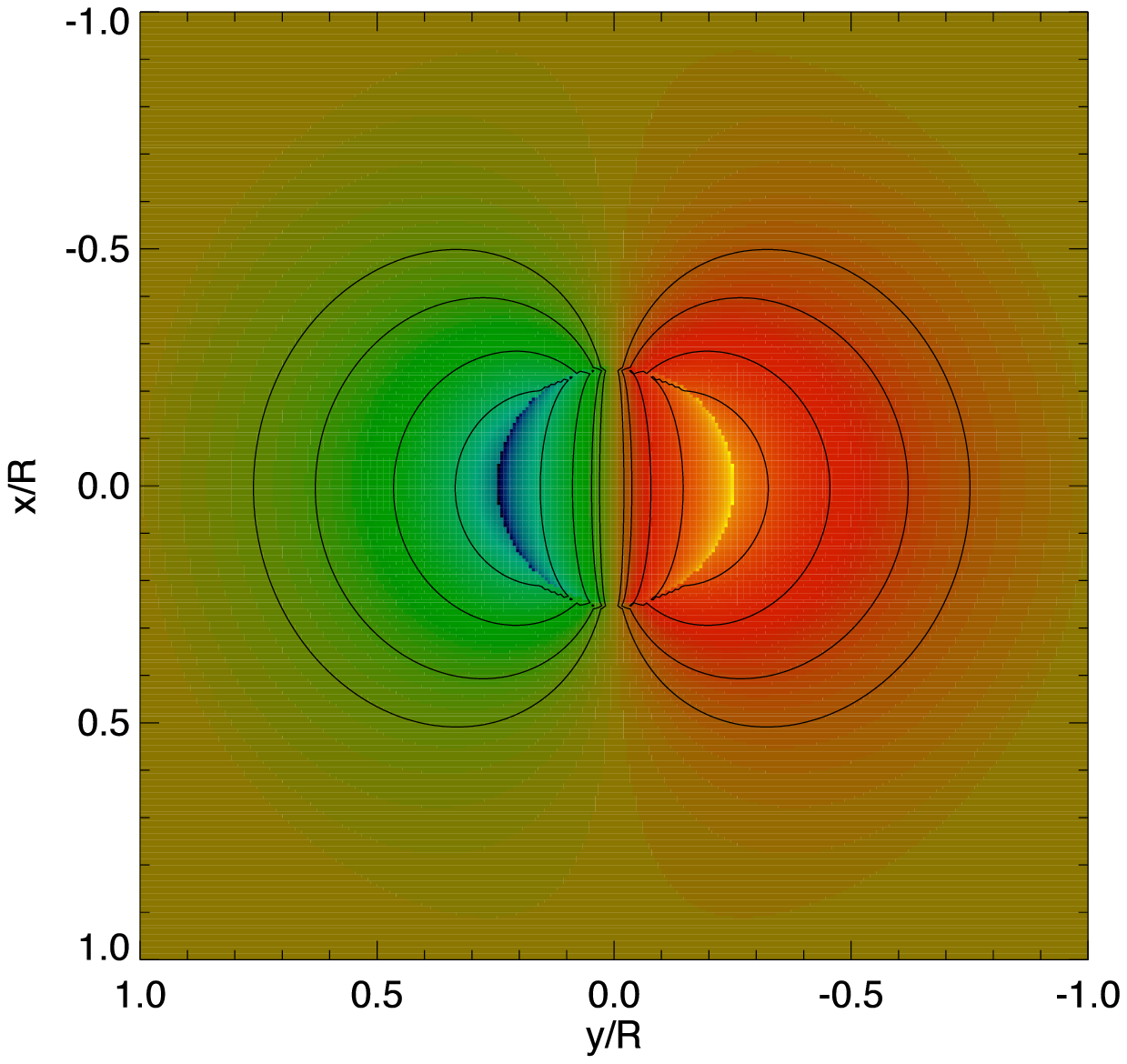}} \\
\centering{\includegraphics[width=3.3in]{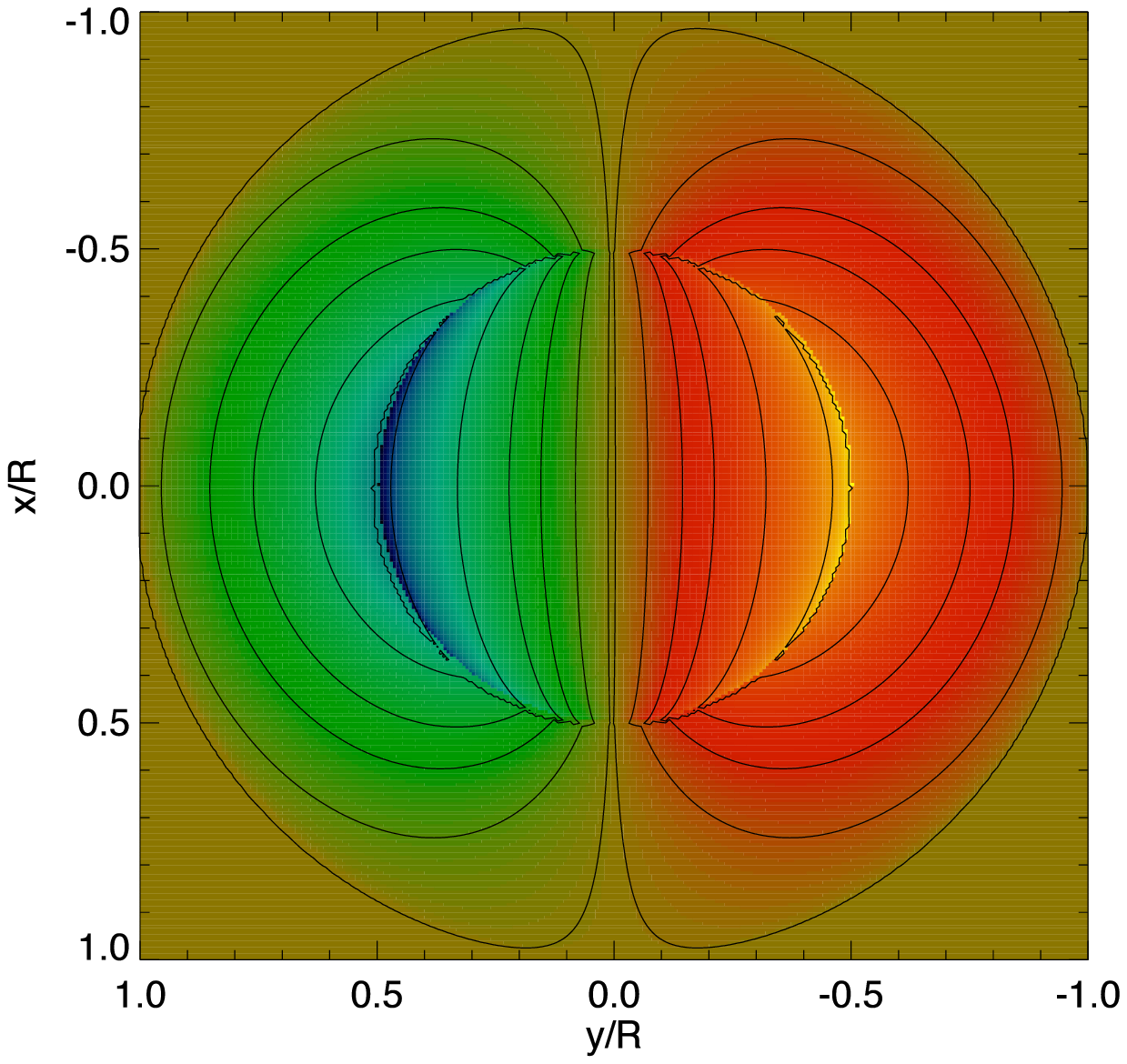}} \\
\centering{\includegraphics[width=3.3in]{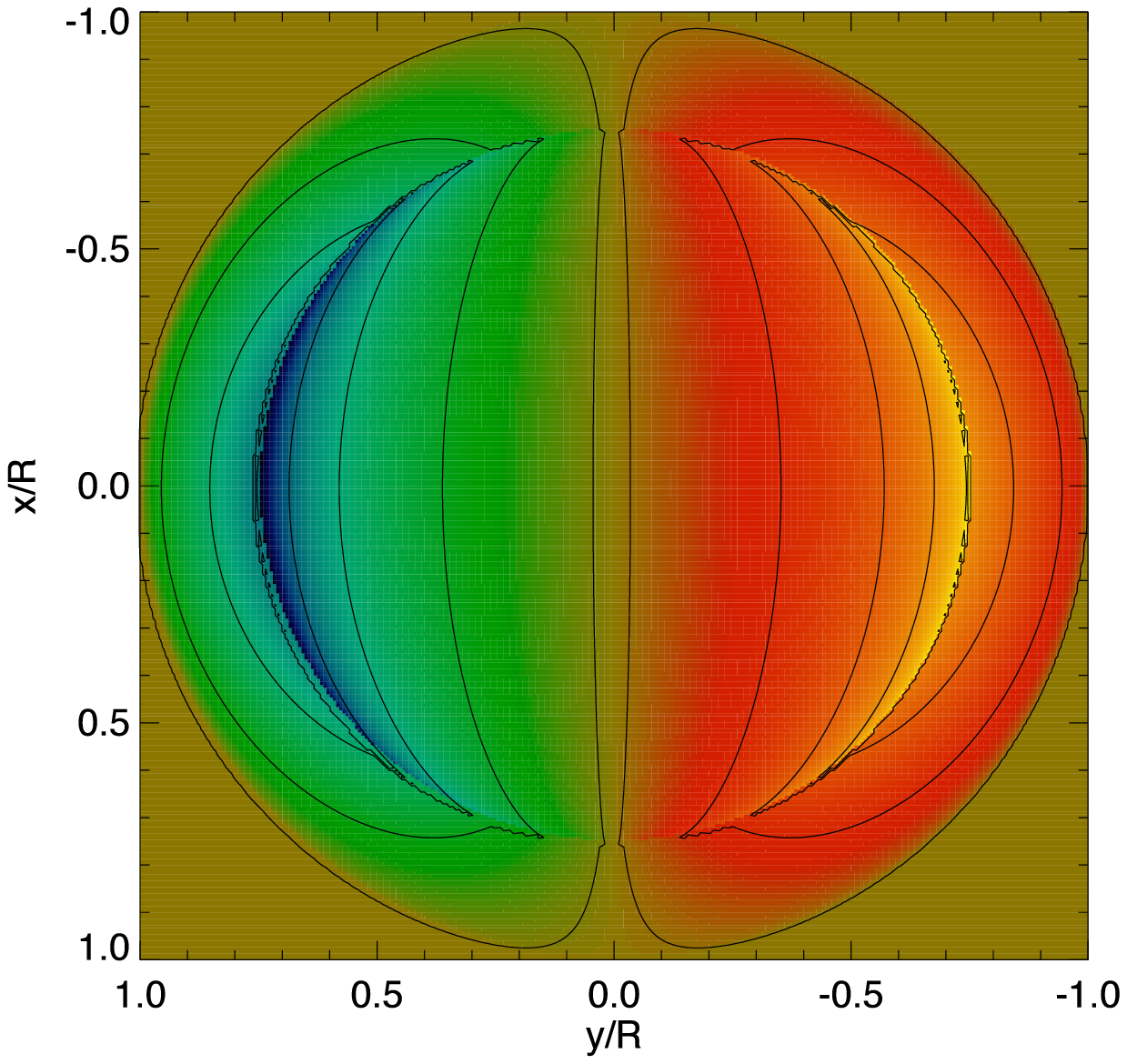}}
\caption{These three panels are for the case of an azimuthal field,
now with an interior spherical region free of any magnetic
field included.  The intent is to approximate a two-wind interaction, 
with an inner, unmagnetized fast wind catching up to an outer slower
and magnetized wind.  The panels, from top to bottom, are for
interior field-free regions with radii of 25\%, 50\%, and 75\% of
the bubble radius.  
\label{fig4}} 
\end{figure}

\begin{figure*}[t!]
\centering{{\includegraphics[width=3.9in]{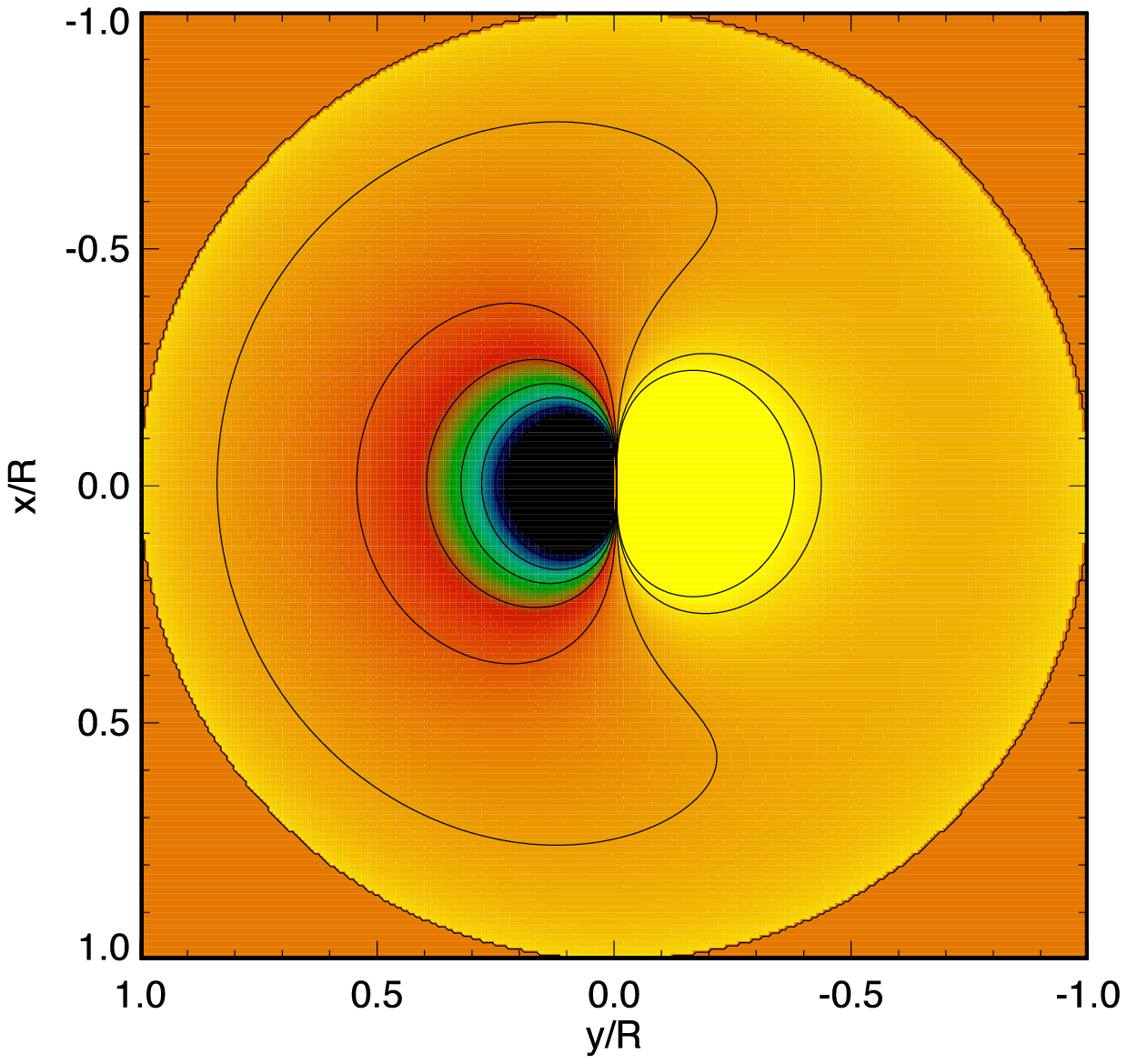}}{\includegraphics[width=3.in]{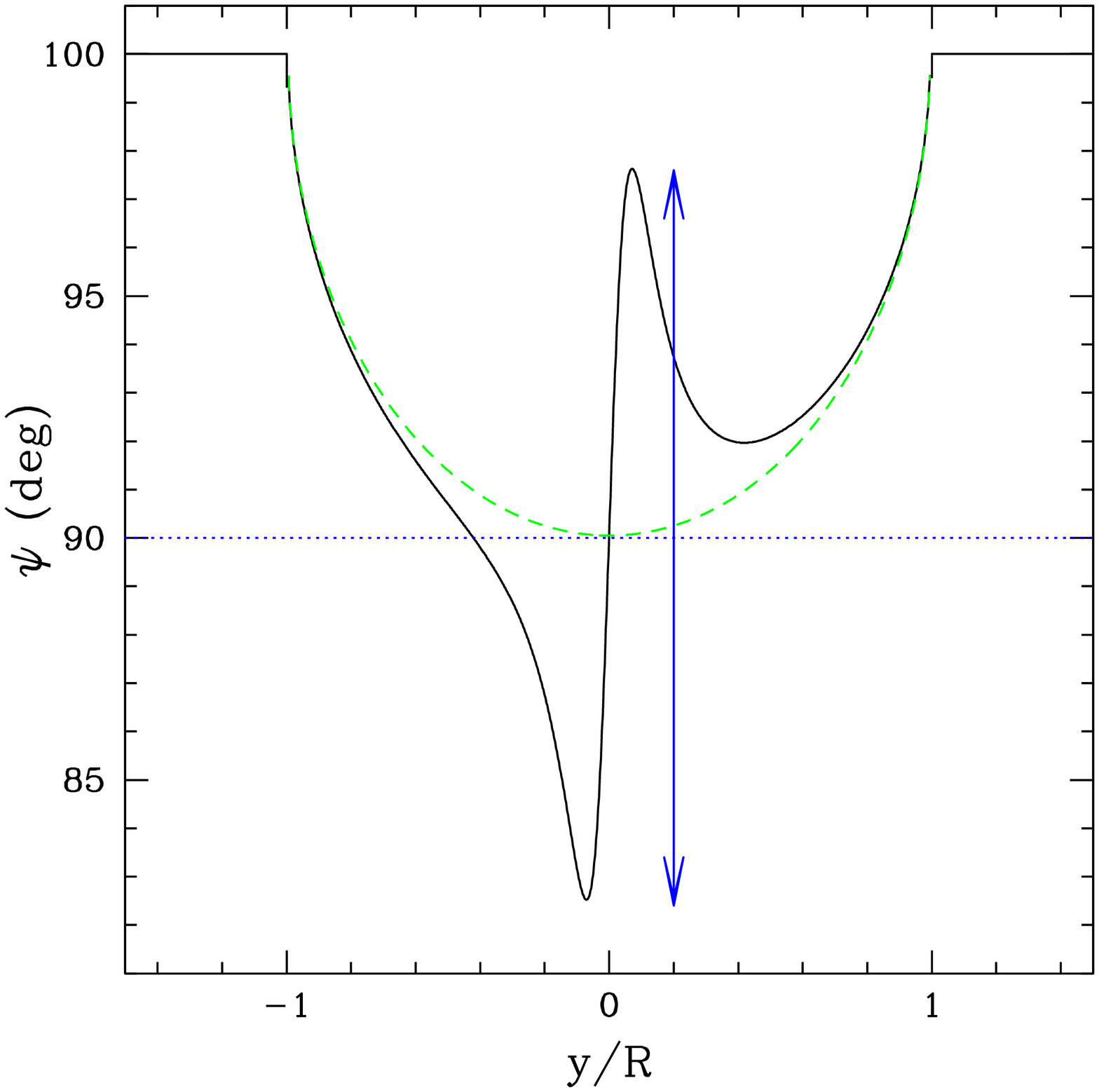}}}
\caption{(a) Left is a PA map for a bubble with an azimuthal field, like
Fig~\ref{fig3}, but now with interstellar polarization effect included
(see text).  (b) Right shows how the polarization PA changes as a function
of $y$ along a horizontal line with $x=0.1R$.  Black is the full profile.  
Green signifies the term $\delta \psi_{ISM}$.  The horizontal blue line is
the half-point between the two local maxima.  Then the vertical blue arrows
represent the full change in PA from one side of the bubble to the other because
of $\psi_{\rm bub}(x,y)$.
\label{fig5}}
\end{figure*}

\begin{figure}[t!]
\plotone{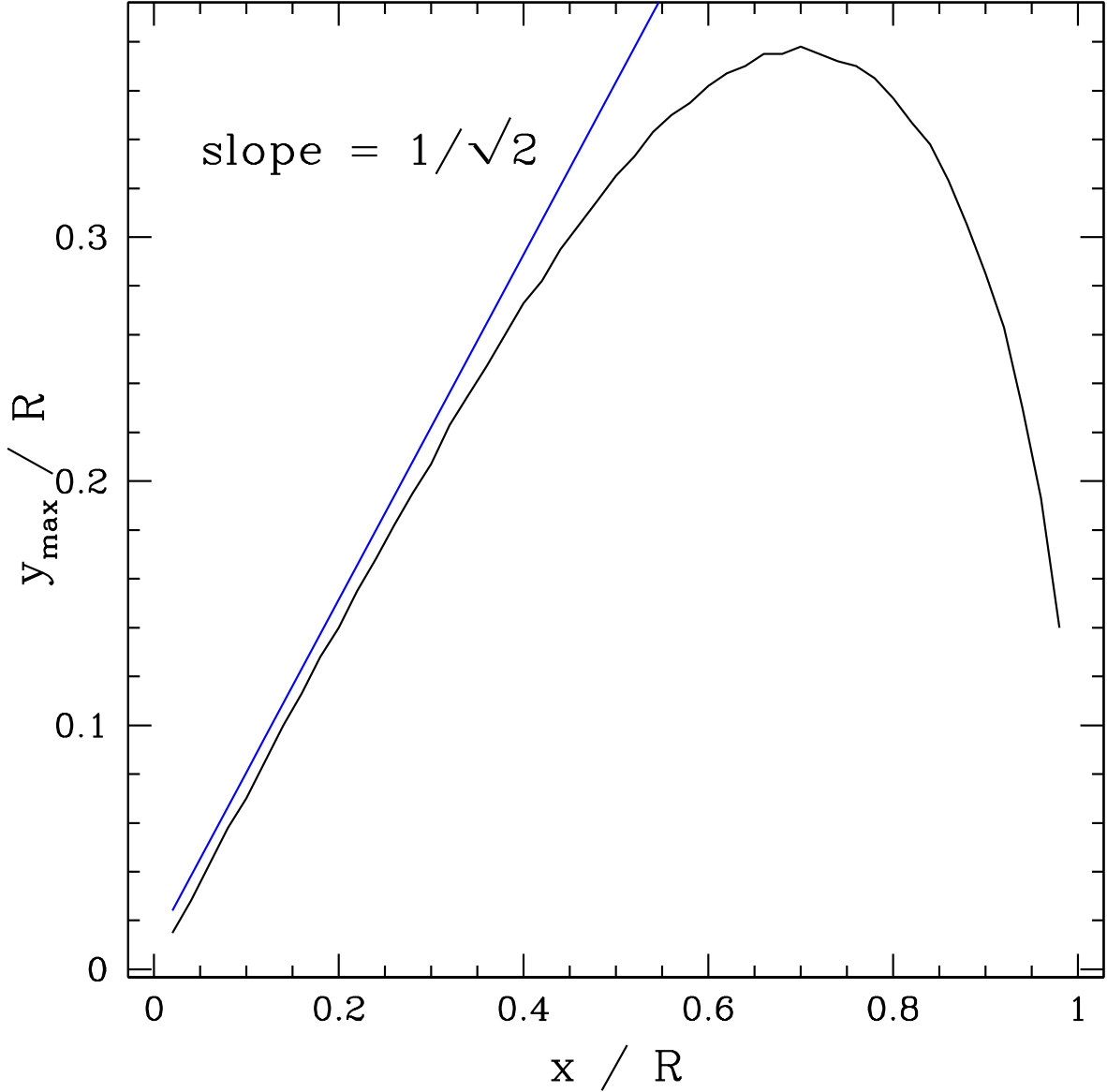}
\caption{Solution for the location of $y_{\rm max}(x)$ where
$\psi_{\rm max}$ is achieved 
in the case of a wind-blown bubble with an azimuthal
magnetic field (but no cavity).  The black line is the numerical
solution.  For $x/R \ll 1$, $y_{\rm max}/R$ is also small, indicating
that $\psi_{\rm max}$ occurs near the projected center of the bubble.
In this limit $y_{\rm max} \approx x/\sqrt{2}$,
as shown by the blue line, which has the correct slope but is
shifted upwards slightly for ease of viewing.
\label{fig6}}
\end{figure}

\begin{figure}
\plotone{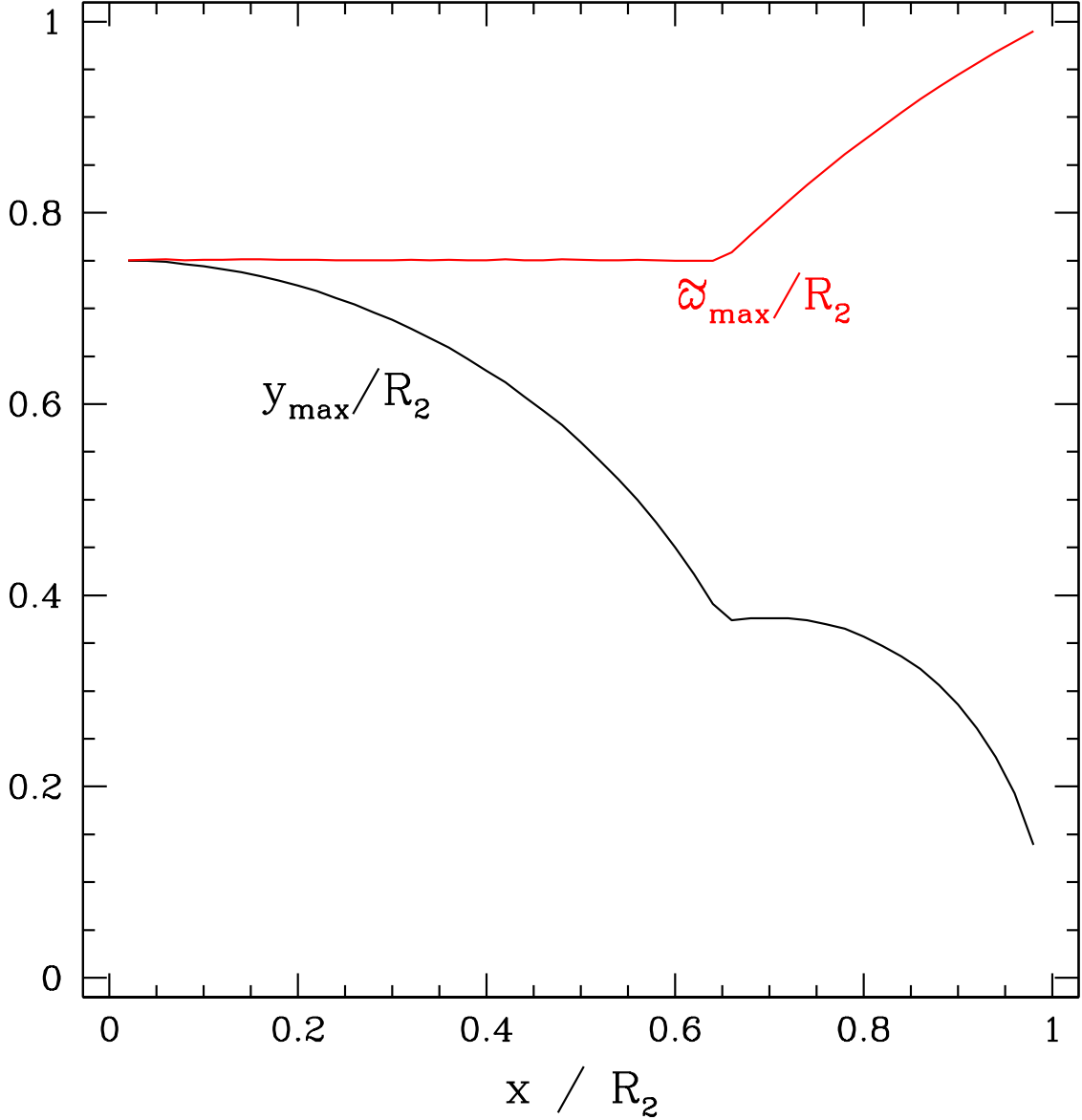}
\caption{Similar to Figure~\ref{fig6}, but now for the case
of a swept-up wind.  In this example the field-free cavity
has a radius $R_1=0.75R_2$.  Solid shows the location
of $y_{\rm max}(x)$.  Also plotted is the
corresponding value $\varpi_{\rm max}=\sqrt{x^2+y_{\rm max}^2}$
in red.  The latter is seen to be flat for $|x|<R_1$
indicating that the location of the peak value $\psi_{\rm max}$ follows 
a circular arc for this range of $x$-values.
\label{fig7}}
\end{figure}

\end{document}